\RequirePackage{ifpdf}
\pdfoutput=1 
\documentclass{JINST}

\usepackage{siunitx}
\usepackage{pdflscape}
\usepackage{gensymb}
\usepackage{graphicx}
\usepackage{lineno}
\usepackage{amsmath}

\title{Scalability study of solid xenon}

\author{J.~Yoo\thanks{Corresponding Author: yoo@fnal.gov}, H.~Cease, W.~F.~Jaskierny, D.~Markley, and R.~B.~Pahlka\\
Fermi National Accelerator Laboratory, Kirk and Pine St., Batavia, IL 60510, USA}
\author{D.~Balakishiyeva and T.~Saab\\
Department of Physics, University of Florida, Gainesville, FL 32611, USA}
\author{M.~Filipenko\\
Erlangen Center for Astroparticle Physics (ECAP), Friedrich Alexander
University of Erlangen-Nuremberg, Erwin-Rommel-Stra{\ss}e 1, 91058
Erlangen, Germany\\
}

\abstract{We report a demonstration of the scalability of optically transparent xenon in the solid phase for use as a particle detector above a kilogram scale. We employed a cryostat cooled by liquid nitrogen combined with a xenon purification and chiller system. A modified {\it Bridgeman's technique} reproduces a large scale optically transparent solid xenon.}

\maketitle

\begin{document}
\section{Introduction}\label{intro}
\par The Standard Model (SM) of particle physics has been explored with amazing accuracies from the scale of the Hubble radius to the size of nucleons. Despite the remarkable success of the SM, a number of observations have recently emerged suggesting the incompleteness of our understanding of fundamental interactions. Three main pieces of evidence contribute to this conclusion: the CP-asymmetry of the Universe, the existence of dark matter and the discovery of massive neutrinos. In all these cases, low background experiments in deep underground sites provide excellent venues to discover new physics Beyond the SM. 

\par Noble elements in both the gas and liquid phases have proven use as excellent low background radiation detectors~\cite{aprile2010,ackerman2011,aprile2012,akerib2014,kamland2013,PhysRevLett.113.121301}. Xenon has drawn special attention among the noble elements due to several distinct advantages over its lighter counterparts. The liquid phase of xenon possesses a very high scintillation light yield (40$\sim$60\, photons/keV) and the vacuum ultraviolet (VUV) wavelength (178\,nm)~\cite{jortner1965} of the scintillation is optically transparent in xenon~\cite{Szydagis:2011tk}. Xenon also has excellent ionization and electron transport properties and the absence of long-lived radioisotopes in xenon results in no intrinsic background radiation sources. The large atomic mass of xenon (131.3 and Z=54) keeps external radioactive decay sources at the outer surfaces of the condensed xenon detector, resulting in improved self-shielding effects. As xenon is a noble element, the chemical purification is straightforward by using hot getter and/or gas distillation systems that can remove most of the non-noble contaminants. A wide range of applications has been studied on particle tracking and spectroscopy including $\gamma$-ray astronomy, neutrinoless double beta decay (NLDBD), dark matter searches, neutrino coherent scattering experiments and medical imaging devices.

\par The solid (crystalline) phase of xenon not only inherits most of these assets from liquid xenon, but also has additional advantages as a particle detector material. Electron drift speeds in thin layers of solid xenon have been measured to be faster compared to those in the liquid phase~\cite{Miller:1968zza}. Optically transparent solid xenon ("crystal xenon") is also transparent to the xenon scintillation lights which are in the VUV range~\cite{jortner1965, Baum:1988, Varding1994, Kubota1982}. Furthermore, an increased amount of light collection in the solid phase compared to the liquid phase using the same detector system in alpha irradiation has been reported~\cite{Aprile1994129}. Therefore, compared to a liquid xenon detector, a solid xenon detector can become an ionization detector with faster response and/or a scintillation detector with improved photo collection. In addition, as the density of solid xenon (3.41 g/cc) is higher than that of liquid xenon (2.95 g/cc), an even further compact detector can be built in the solid phase. One may also imagine a low energy bolometric phonon readout from solid xenon with superb energy resolution. Therefore, solid xenon can be a strong contender of low background and low energy threshold dark matter detector. Solid xenon can also be a great NLDBD detector material, where the Ba$^{++}$ ions that are produced from the xenon NLDBD would be frozen at their decay locations and subsequently identified via ion-tagging method. The double coincidence (energy and ion-tagging) will substantially reduce the NLDBD background in the energy region of interest. 

\par Xenon in the solid phase is chemically stable and forms simple face-centered cubic crystal structures with interatomic binding energies coming from weak Van der Waals forces{\footnote {The conditions imposed by mechanical stability limit the possible structures of ideal xenon crystals. These are generally face-centered-cubic (fcc) or hexagonal-close-packing (hcp). The crystal structure of xenon changes fcc to hcp under pressure above $\sim$20 GPa~\cite{Caldwell15081997}. The lattice cell parameter is 6.2~\AA~in fcc.}}. It is known to be relatively easy to produce small-scale clear and transparent solid noble element specimens that are virtually perfect crystals{\footnote{These crystals nevertheless are polycrystalline in many cases and contains a large number of microscopic defects. The density of structural defects is highly depends on the growing conditions (temperature and pressure) of the crystal. For example, noble element crystals grown under vapor deposition method shows density of dislocations of order 10$^6$ cm$^{-2}$, while commercial silicon wafers shows dislocation density of order 10$\sim$1000 cm$^{-2}$ depends on the methods of fabrication. For more details about the microscopic properties of noble gas solids, see reference~\cite{RGSv2}.}~\cite{RGSv2}. The crystal structure~\cite{Venables1972, Kramer1972, Kramer1976, Niebel1974} and microscopic defects~\cite{Venables1966} are important properties for experimental applications such as solar axion searches{\footnote{The directional dark matter searches with crystal detectors would be a challenge due to the short ($\sim\mu$m of) track length of the recoil nucleus.}~\cite{Ahmed:2009ht}. However, for most particle detector applications, it is important to first understand the macroscopic properties such as density uniformity, optical transparency, charge drift velocity, scintillation, and ionization. Particle detectors based on the solid phase of noble element active media have been thoroughly investigated~\cite{Miller:1968zza, Aprile1994129, Bolozdynya1977, Himi1982, Kubota1982, Kink1987, Bald1962,  Varding1994, Michniak2002387, Gushchin1982, Aprile:1985xz}. Even though most of these studies have successfully shown that the solid noble elements are excellent candidates for particle detector material, large scale detectors have yet to be realized. Our first R\&D effort, therefore, is focused on the proof of scaleability of optically transparent solid xenon that is distinguishable from opaque frozen xenon bulk. In this paper, we describe the instrument details and demonstrate the scalability of solid xenon above a kilogram scale.

\section{Solid xenon test stand}\label{sec:teststand}
\par Figure~\ref{fig:sxsystem} shows the schematics of the solid xenon cryostat. The cryostat consists of a stainless steel vacuum jacketed chamber with an outermost diameter of 30\,cm with three 15\,cm diameter glass window ports. Two concentrically placed glass chambers reside inside allowing optical access to the xenon bulk volume.  The larger of the two glass chambers is used as a liquid nitrogen bath for cooling and has a diameter of 23\,cm, referred to as the LN (liquid nitrogen) chamber. The smaller 10\,cm diameter inner chamber houses the xenon volume and is made out of Pyrex with a 5\,mm thick side wall and a 10\,mm thick flat bottom, referred to as the xenon chamber. The stainless steel chamber also functions as a safety protection chamber in case of unexpected pressure changes in the glass chambers. The LN chamber is equipped with an effective phase separator at the bottom, made with a combination of aluminum and polyethylene blocks. The phase separator, when it is pre-cooled to liquid nitrogen temperature, transfers nitrogen liquid into the LN chamber with minimal evaporation in the internal transfer line. The turbo line on the vacuum side is used for active vacuum shielding of the cryogenic chamber, and the turbo line on the xenon chamber is used to evacuate the chamber before the xenon transfer. The system has safety backup system which includes a 440 liter xenon recovery cylinder, a 250 liter buffer tank, two 4 liter xenon storage cylinders and a 4 liter cylinder for calibration gas. A commercial hot getter (PF4-C3-R-1 and Monotorr PS4-MT3-R-1 by SAES) and a circulation loop allows continuous purification of the xenon.

\begin{figure}[!t]
\begin{center}
	\includegraphics[width=0.8\columnwidth]{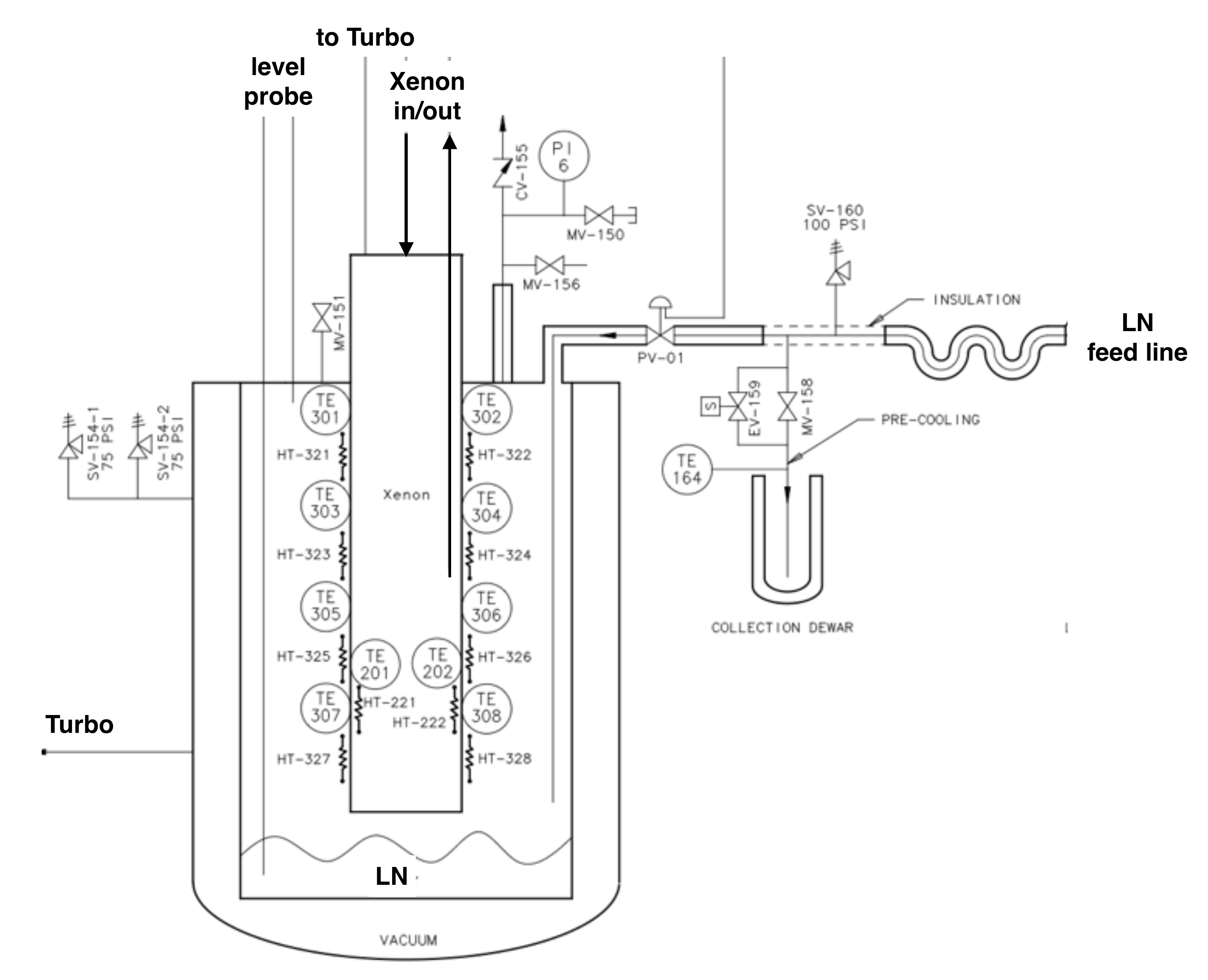}
	\caption{A schematic diagram of the solid xenon test stand~\label{fig:sxsystem}}
\end{center}
\end{figure}

\par For the temperature control of the inner chamber, we employ ten Nichrome 32-gauge heater wires and platinum Resistance Temperature Detectors (RTD: PT-100) that are affixed to the bottom and barrel surface of the xenon chamber using cryogenic epoxy. A total of nine RTDs measure the thermal gradients of the outer surface of the xenon chamber and one thermometer-heater loop is installed near the bottom of the inner wall of the xenon chamber. While the cryogenic features of the system reported here largely overlap with those of liquid-based systems, we require additional advanced controls for temperature and pressure in order to keep the optical transparency without obvious defects in the solid xenon. The glass chambers place constraints on the pressure in the inner glass chamber, set to about 1\,bar. 

\par A main control panel is designed to handle the xenon transfer and monitoring at a centralized location. It allows convenient monitoring and controlling of parameters, such as the xenon flow rate, cooling bath level, temperatures, pressures, and weights of the xenon storage cylinders. The cooling LN is fed from a continuously-serviced 20 ton external LN tank that facilitates uninterrupted cryogenic system operation. Using the cold nitrogen gas above the liquid to cool the xenon chamber, the liquid nitrogen level in the LN chamber is controlled by a feed-back system of pneumatic control valves and differential pressure level meters with an accuracy of about 0.5\,cm. The liquid nitrogen level is set to about 2\,cm below the bottom of the inner glass chamber during normal operation and is therefore not in direct contact with the xenon chamber. The temperature of the glass chamber is controlled using three sets of thermometer--heater feedback loops (bottom, barrel, and top). The insulation vacuum between the stainless steel chamber and the LN chamber is maintained at about $10^{-5}$\,Torr using a dedicated turbo pump. 

\par The xenon chamber is cleaned in a ultrasonic bath before the installation and is then baked at 40\,\degree C for a few days using the attached heaters. The vacuum level of the xenon vessel reaches $10^{-8}$\,Torr at room temperature with no detector instruments in the chamber and stabilizes at this value at a temperature of 164\,K.  With detector instruments installed in the chamber, the vacuum level reaches below $10^{-5}$\,Torr after two days of evacuation. We used research grade xenon with 99.999\% purity level. The other components reported by the gas provider are: krypton ($<$ 1 ppm), water ($<$ 1 ppm), hydrogen ($<$ 0.5 ppm), oxygen ($<$ 1 ppm), nitrogen ($<$ 2 ppm), hydrocarbons ($<$ 1 ppm), tetrafluoromethane ($<$ 1 ppm), and carbon dioxide ($<$ 0.5 ppm). 
These contaminations are all below the capable range of our Residual Gas Analyzer (RGA) ($<$ 100 ppm). No gas contamination above the RGA background level is observed in the xenon gas.

\section{Scalability of solid xenon}\label{sec:scalability}
\par The main focus of the scalability study is to understand the conditions required to produce optically transparent solid-phase xenon which is distinguishable from frozen opaque volumes or other types of solid phases. Due to the density difference between the liquid phase (2.95 g/cc) and the solid phase (3.41 g/cc), the growing process of solid xenon requires special care in maintaining and controlling the growing speed.  

\par The configuration of the heater wires on the inner glass chamber is designed to produce 2\,kg of solid xenon at maximum. We found that a transparent layer of solid xenon can be reliably grown via vapor deposition methods in a pre-cooled chamber. However, the growth rate of solid xenon was quite slow ($\sim$5\,mm/day initially, then gradually slows down), and the shapes of the solid surfaces were non-uniform. We concluded that the vapor deposition method would be more appropriate for thin layer detectors rather than for large scale detectors. 

\par A modified {\it Bridgeman's technique}~\cite{RGSv2} was adopted to grow solid xenon on a kg-scale. In Bridgeman's method, a temperature gradient of 1$\sim$2\,K/cm is established across the liquid, then growing from the liquid to solid at about 1\,bar of pressure, which is then cooled progressively from the bottom at a rate of $\sim$1\,K/hr to allow the heat from solidification to dissipate, and for annealing to take place.

\begin{figure}[t!]
\begin{center}
\includegraphics[width=5in]{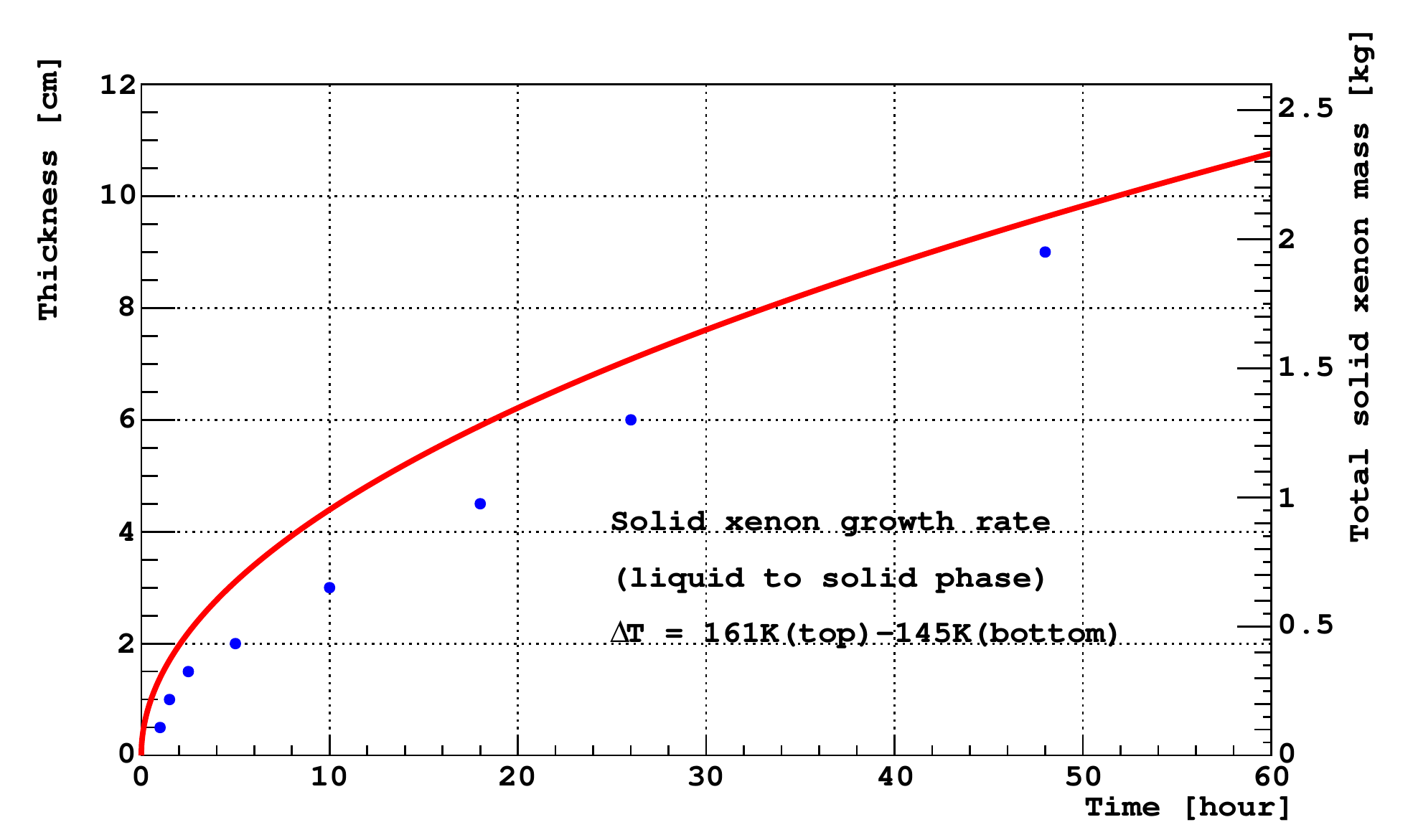}
\caption{A simple thermal model of the expected solid xenon growth (red curve), and an example of actual growth of solid xenon (blue dots). The temperatures at the bottom and top of the liquid xenon volume are set to 145 and 161 K, respectively. The disagreement between the ideal 1-dimensional thermal model (red curve) and actual growing speed (blue dots) might be due to different thermal configuration of the realistic xenon glass cylinder and the local convection effect of the liquid xenon above the solid xenon.~\label{fig:sxgrow}}
\end{center}
\end{figure}

\begin{figure}[t!]
\begin{center}
\includegraphics[height=2.in]{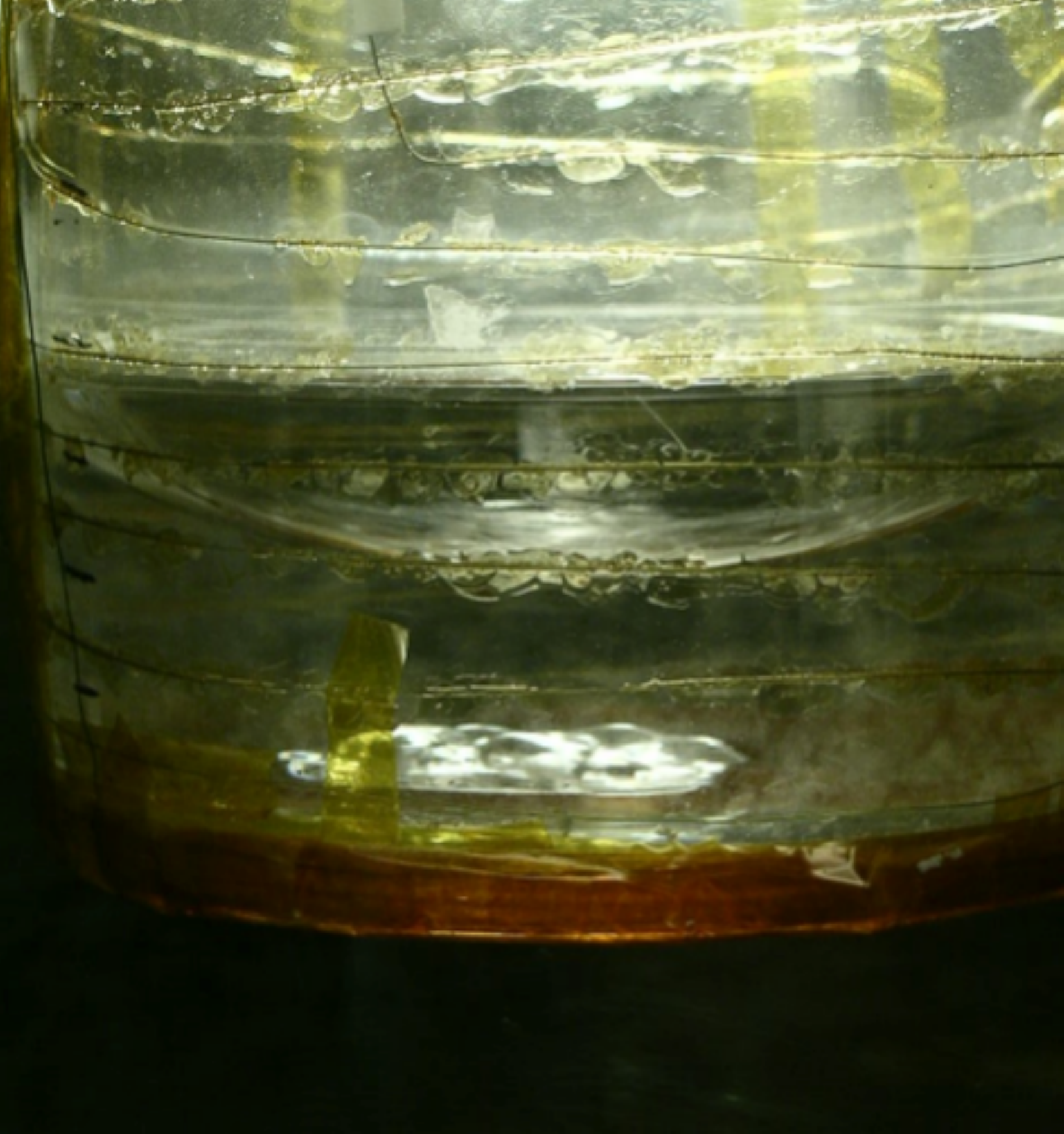}
\includegraphics[height=2.in]{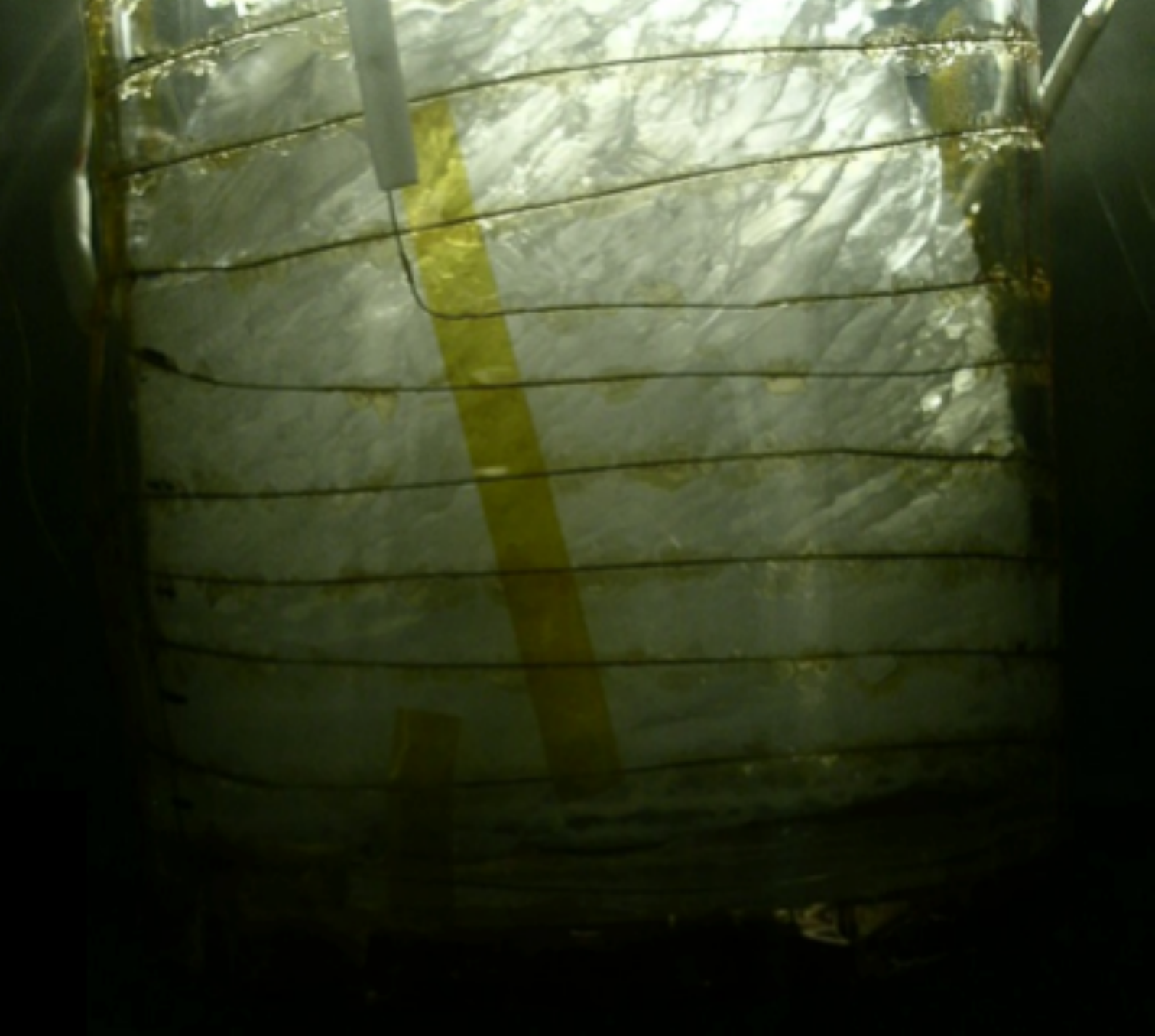}
\includegraphics[width=4.15in]{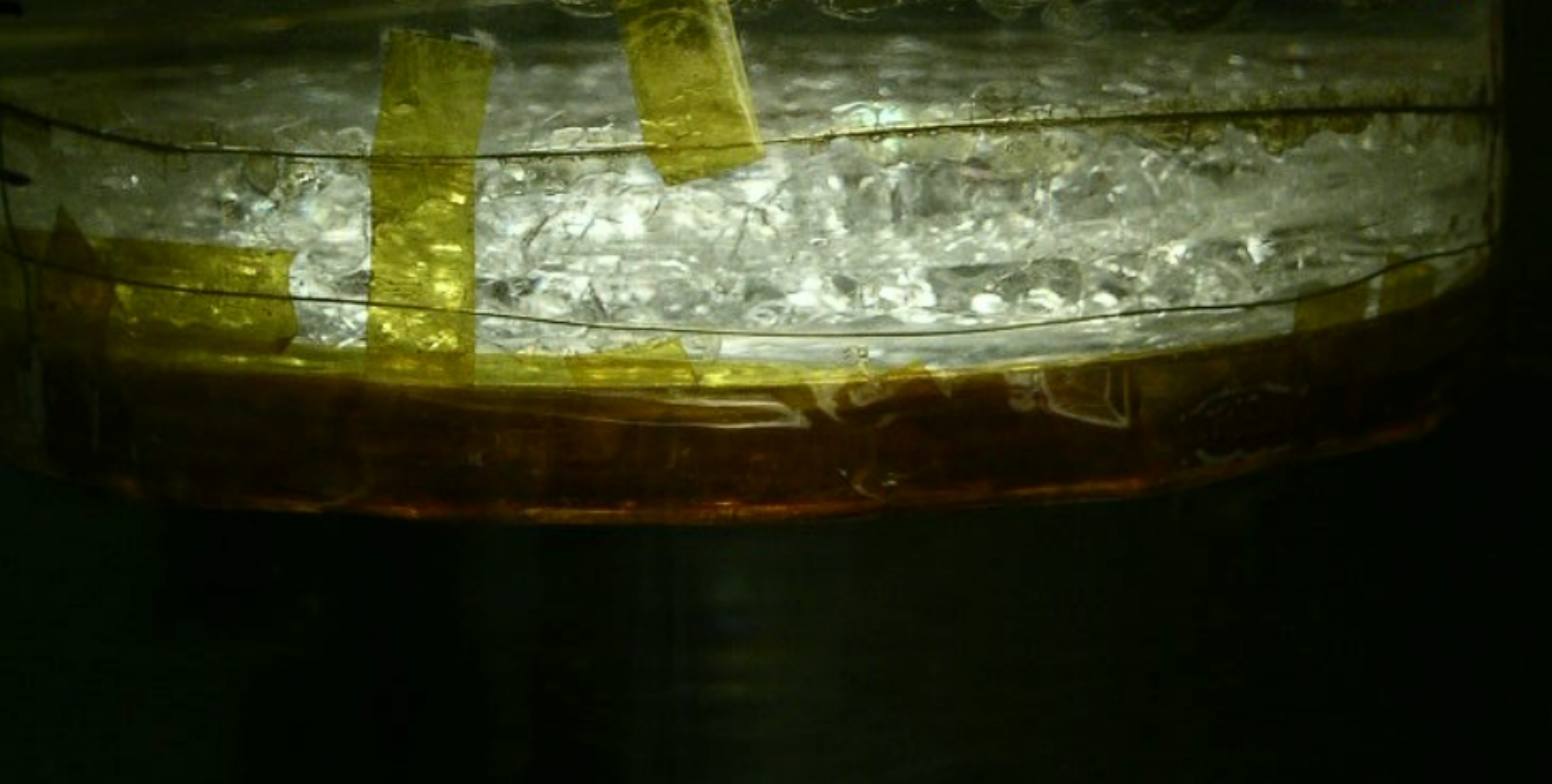}
\caption{Photographs of various types of solid xenon. The top left photo shows a concave structure of the solid surface due to the over-cooling at the barrel. There is also an opaque substructure at the bottom center due to the significant temperature gradient at the bottom surface. These are produced when there was a rapid variation of cooling liquid level. The top right photo shows frozen layers of solid xenon. These structures can be easily created when the pressure drops below the vapor pressure. The bottom photo shows solid xenon in a polycrystalline state on the order of a few mm, provided when the temperature and pressure control become unstable near the triple point (161.4\,K and 11.9 PSIA). The dark brown color at the bottom of the chamber is kapton tape. The spacing of the heater between each wire is about 1\,cm, which is a good measure of the approximate height of xenon.~\label{fig:frozenxenon}}
\end{center}
\end{figure}

\par Due to the slow growth rate of the transparent solid phase of xenon, the bottom-to-top temperature gradient requires tuning that depends on the size of the solid xenon volume. In ideal conditions, the growth rate of solid xenon from the liquid phase can be estimated by $L = \sqrt{(2K~\Delta T)/(H\rho)} \sqrt{t}$, where $L$ is the thickness, $t$ is time, the latent heat of xenon fusion is $H$=17.5\,J/g, the thermal conductivity of solid xenon is $K$=0.001\,W/(cm$\cdot$K)~\cite{Purskii2004}, and the solid xenon density is $\rho$=3.41/cc. Figure~\ref{fig:sxgrow} shows a simple thermal model of the expected solid xenon growth rate (red curve) and an example of actual growth data (blue dots). In this example, the bottom temperature is set to 145\,K and top temperature is set to 161\,K. The actual growth rate of solid xenon is a little slower than expected by the simple model. In order to grow 2\,kg of solid xenon, which is equivalent about 9\,cm high in the xenon chamber, it requires more than 48 hours of stable temperature control. The details of the thermal configuration in the solid xenon chamber system requires fine tuning of the temperature gradient. We found the following setup reliably reproduces the optically transparent solid xenon in our system. 

\par First, the xenon is liquified in the glass chamber at a set temperature of 163\,K and pressure of 14.5$\pm$0.5\,PSIA. When the liquid xenon level reached about 4.5\,cm height in the xenon chamber, the bottom of the chamber was slowly cooled down to 145$\pm$0.5\,K, while cooling the barrel part of the chamber to 157$\pm$0.5\,K. The thick glass bottom (1\,cm) and barrel wall (0.5\,cm) reduce the thermal shock at the inner surface of the glass. Once these temperatures are set, the solid xenon forms at the bottom of the liquid xenon with an initial growth rate of about a cm per hour, which gradually slows down. Under these conditions, the solid xenon grows almost uniformly over the 9\,cm diameter chamber. We found the temperature balance at the barrel and bottom of the glass chamber is important for uniform growth of the solid surface. If the barrel temperature is too low (150\,K or lower), the growing speed at the barrel exceeds that of the center and begins to form a concave solid surface. Any substantial temperature change produces optical defects in the solid xenon; including opaque spots, filamentary structures and voids. We do not observe any significant pressure dependence of the optical quality of the solid xenon so long as the pressure is kept well above the vapor pressure. Rapid pressure variations near the vapor pressure can easily produce opaque layers at the top surface of the solid volume. Failure to maintain temperature and pressure stability near the triple point creates a few millimeters of polycrystalline xenon which is clearly identified by the birefringence of each polycrystalline cell. Figure~\ref{fig:frozenxenon} shows a few examples of the solid phases of xenon which were produced during the initial test of the solid xenon cryogenics.

\begin{figure}[t!]
\begin{center}
\includegraphics[height=2.3in]{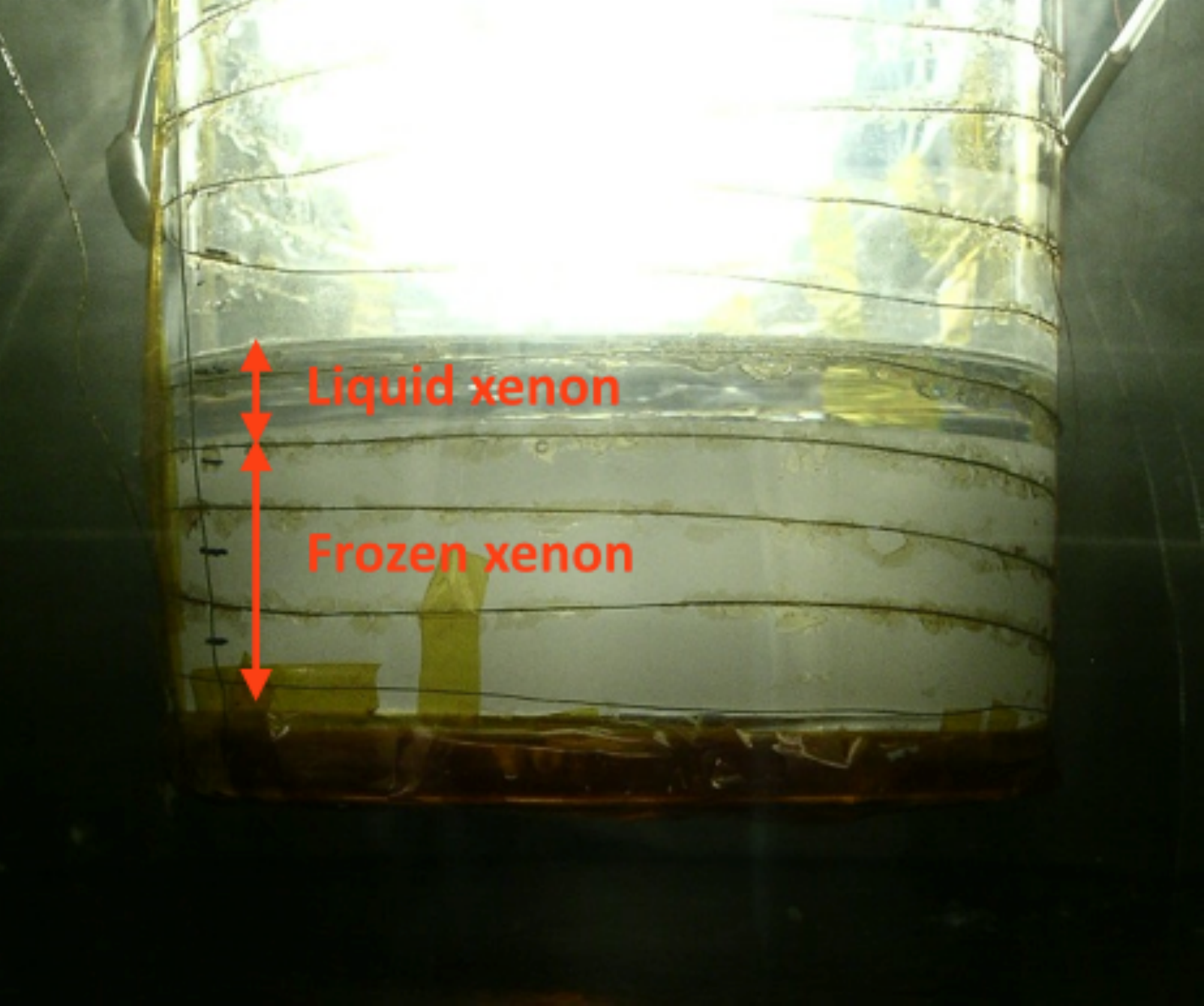}
\includegraphics[height=2.3in]{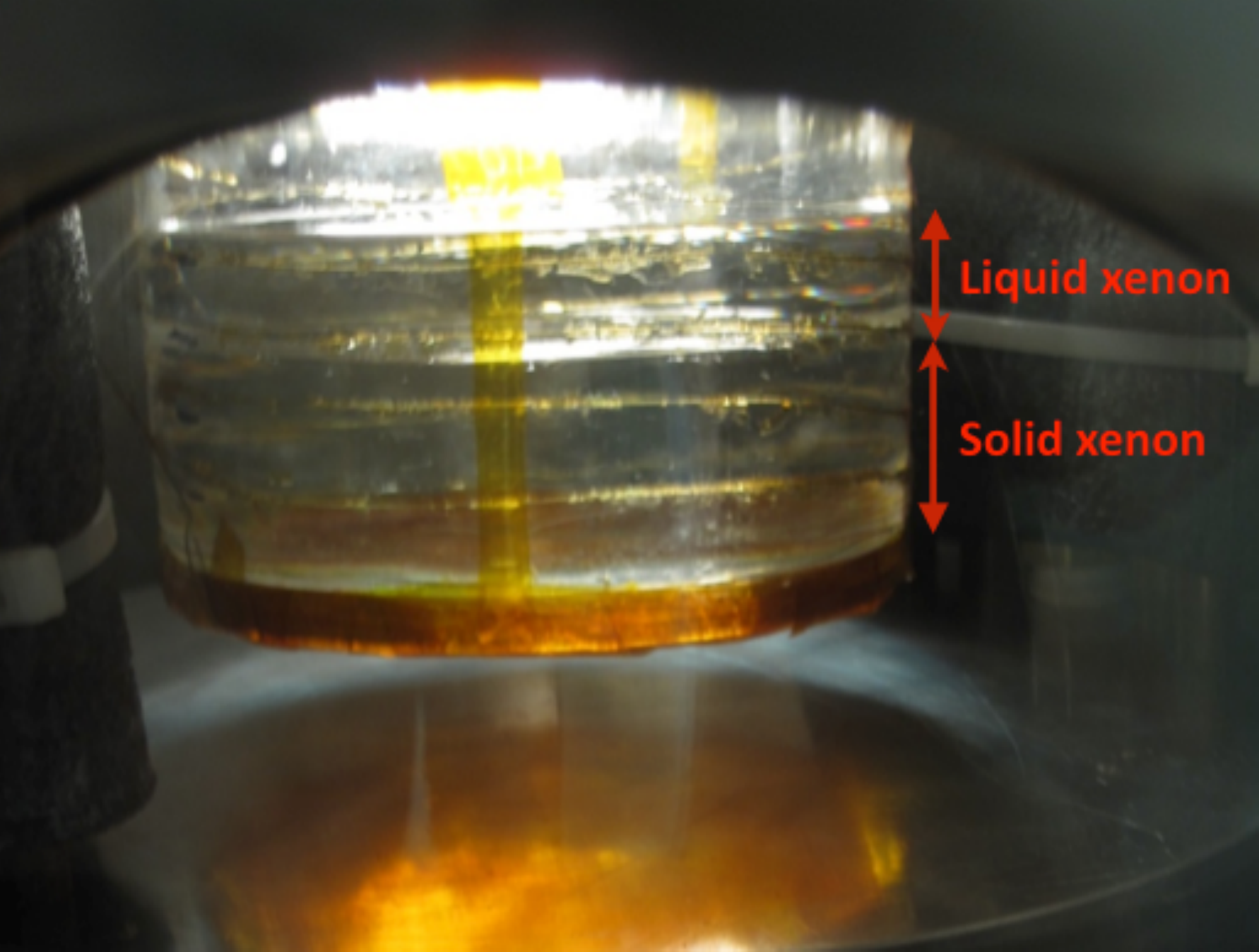}
\caption{Photographs of opaque solid xenon (left) and transparent solid xenon (right). The left photograph shows about 3\,cm of opaque solid xenon and 1\,cm of liquid xenon above the solid phase. The right photograph shows about 3\,cm high (about 650\,g) of transparent solid xenon and 1\,cm (about 189.4\,g) high  liquid xenon on top of the solid xenon.\label{fig:solidxenon}}
\end{center}
\end{figure}

\par Figure~\ref{fig:solidxenon} shows an example of the clear visual difference between frozen opaque xenon (left) and transparent solid xenon (right). Opaque xenon is produced when the temperature at the bottom of the glass chamber was reduced quickly ($\Delta T$/$\Delta t$<$\sim$3\,K/min). The transparent area above the opaque volume in the left photograph depicts xenon in the liquid phase. The right photograph in Figure~\ref{fig:solidxenon} shows about a 3\,cm-thick sample of solid xenon at the bottom of the chamber and about 1\,cm-thick liquid xenon sample on top of the solid xenon. In this case, the boundaries between the liquid phase and the solid phase can be seen by the reflection of light at the surface of the solid volume or a change of refractive index at the boundary between liquid and solid. 

Once the solid xenon is grown at the desired height we slowly raise the bottom temperature to the desired temperature, which is normally 157\,K (or 159\,K) while keeping the barrel temperature at 157\,K (or 159\,K). The largest volume of transparent solid xenon produced in the current solid xenon chamber was about 2\,kg at a temperature of 157\,K (see Figure~\ref{fig:sxetpc}, for example). The temperature of the large scale solid xenon can be further reduced by a very slow control of temperature ($\Delta$T$\simeq$1\,K/day), however, we occasionally observed fuzzy opaque spots and structural defects in the solid volume at the temperature below 155\,K. It took more than two days starting from liquid phase to grow a cylinder of solid xenon that is 9\,cm in diameter and 9\,cm high. The transparency can be kept for an extended period of time under stable temperatures and pressures. The longest duration where the solid xenon was kept defect-free and transparent was 10-days until the test discontinued due to other maintenance issues of the experimental building. The temperature and pressure are the only two parameters that are related to maintain the optical transparency of the solid xenon volume for an extended period of time. However, there is a possibility that high energy cosmic-rays may leave visible structural damage in the solid xenon, though we have not observed any visible changes during our tests.

\begin{figure}[t!]
\begin{center}
\includegraphics[width=2.8in]{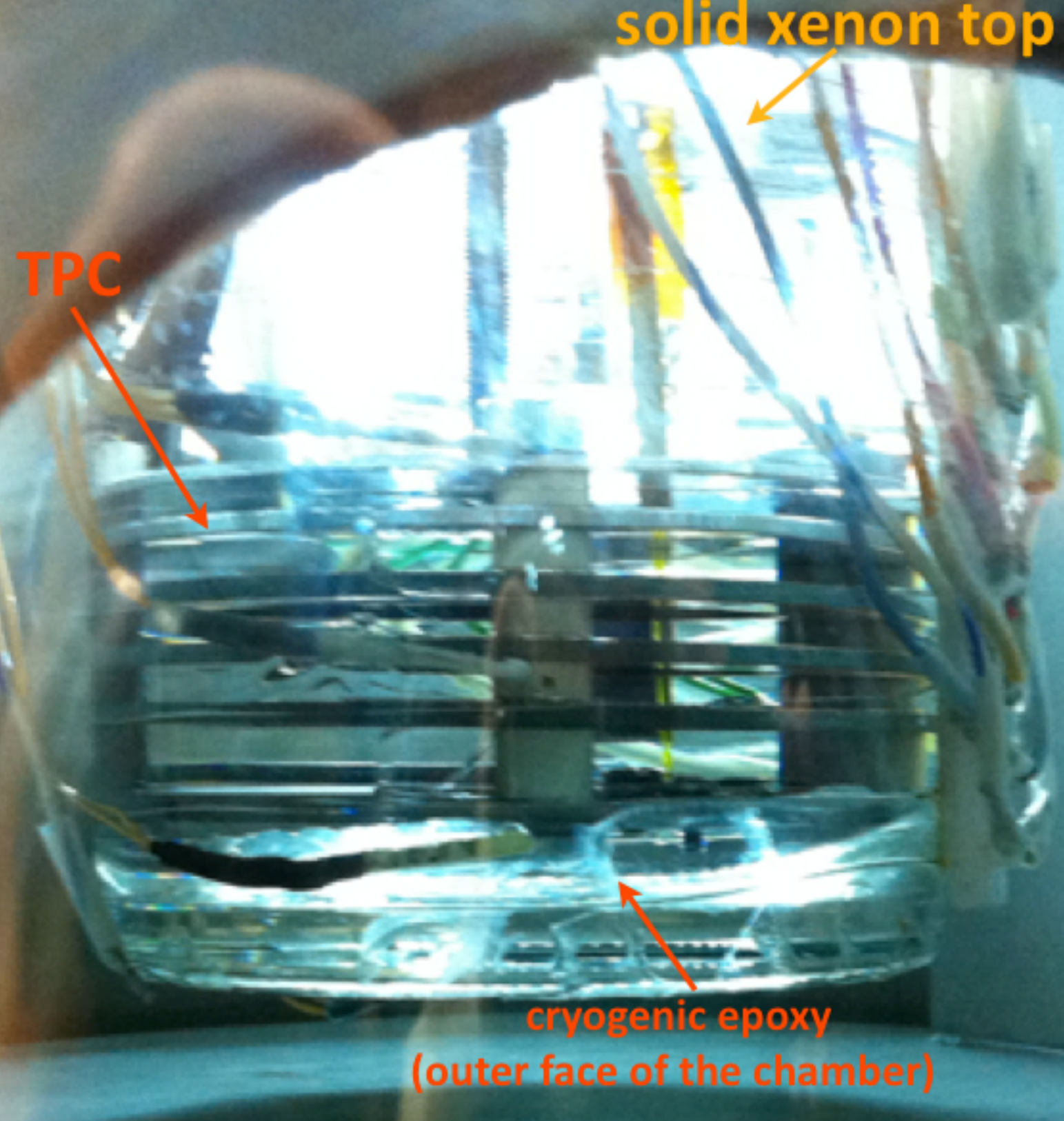}
\caption{The time projection chamber (TPC) installed in the xenon glass chamber. In this particular sample, the amount of solid xenon in the chamber is about $\sim$2\,kg. A little opaque area near the surface of the glass bottom edge is cryogenic epoxy. In this particular sample, some birefringence can be seen at the inner glass wall near the grid rings.~\label{fig:sxetpc}}
\end{center}
\end{figure}

\section{Discussion and Summary}\label{sec:discussion}
\par We demonstrated the scalability of transparent solid xenon above a kg-scale. We found a modified {\it Bridgeman's technique} is suitable to reproduce optically transparent solid xenon in a consistent manner. The solid xenon production is automatically controlled using heater and thermometer loops in a gas-cooled glass chamber system. Our system is designed to maximize visual access to the xenon volume while keeping the glass chamber in a safe condition. We produced about 2\,kg of optically transparent solid xenon at a temperature of 157\,K. To our knowledge, this is the largest single monolithic volume of optically transparent solid xenon that have been produced and reported. Developing solid xenon larger than a few kg would require substantial modifications to our existing system as the current set up may not be the best design to achieve a uniform thermal configuration for a larger scale solid xenon. A competitive dark matter experiment would require a ton-scale background free detector. However, it would take more than 200 days to grow solid xenon to a height of 1 m using the same method that we used with a similar thermal configuration. Therefore, a substantial improvement in solidifying speed is required to develop a ton-scale detector. For example, a cold-bath method, where the xenon chamber slowly sinks into a temperature-calibrated cryogenic liquid bath, would be an attractive alternative when building a larger scale detector.

\par The next interesting R\&D topics are understanding the scintillation and charge transport properties in a kg scale of solid xenon. Bolometric readout of energy deposit in heat channel would be quite an interesting research topic in the future and will open up the possibility of measuring all three energy deposition channels of particle interactions -- scintillation, ionization and heat. However developing solidification process of large scale crystal xenon at sub-Kelvin temperature appears technically challenging. The vapor deposition method would be a good test option for a low temperature and a thin-layer (less than a cm scale) of solid xenon detector development. We can consistently reproduce transparent solid xenon bulk that was 5\,mm thick at the temperature of 77\,K via the vapor deposition method.   

\par In summary, the research and development efforts towards employing solid xenon as a particle detector were presented. Using a three-chamber vessel with automated liquid nitrogen cooling conditions, we demonstrated the scalability of optically transparent solid xenon above a kilogram scale.

\acknowledgments
We are very grateful to M.~Miyajima, J.~White, and A.~Bolozdnya for the initial discussions of the solid xenon particle detector and sharing their ideas. We thank R.~Barger, D.~Butler, R.~Davis, A.~Lathrop, L.~Harbacek, K.~Hardin, C.~Kendziora, W.~Miner, K.~Taheri,  M.~Rushmann, E.~Skup, M.~Sarychev and J.~Vorin at Fermilab for their tireless hard work to provide us the experimental setup with highest standard. We also thank V.~Anjur, A.~Anton and B.~Loer for their participation of the system setup. This work supported by the Department Of Energy Advanced Detector R\&D funding.

\bibliographystyle{h-physrev3}
\bibliography{sxref}{}
\end{document}